\documentclass[aps,prd,preprint,nofootinbib]{revtex4-1}
\usepackage{graphicx}
\usepackage{braket}
\usepackage{amsmath}
\usepackage{amssymb}
\usepackage{dsfont}
\usepackage{hyperref}
\usepackage{interval}
\usepackage{bm}
\usepackage{picture}
\usepackage{cancel}
\usepackage{subcaption}
\usepackage[x11names]{xcolor}
\usepackage{csvsimple}

\def\draft{0}

\if\draft1
\usepackage[markup=underlined]{changes}
\else
\usepackage[final]{changes}
\fi

\newcommand{\stkout}[1]{\ifmmode\text{\sout{\ensuremath{#1}}}\else\sout{#1}\fi}
\setdeletedmarkup{\stkout{#1}}

\intervalconfig {
	soft open fences ,
}
\DeclareMathOperator{\tr}{tr}
\DeclareMathOperator{\lspan}{span}
\DeclareMathOperator{\Cl}{Cl}
\DeclareMathOperator{\diag}{diag}
\DeclareMathOperator{\dg}{d}
\newcommand{\HS}{\mathcal{H}}
\newcommand{\HSu}{\mathcal{H}^{\text{unphys.}}}
\newcommand{\HSp}{\mathcal{H}^{\text{phys.}}}

\newcommand{\Al}{\mathcal{A}}
\newcommand{\dof}{\emph{d.o.f.}}
\newcommand{\GT}{\mathcal{G}}
\newcommand{\MC}{\mathcal{C}}
\def\phys{^{\text{phys.}}}
\def\unphys{^{\text{unphys.}}}

\newcommand*\diff{\mathop{}\!\mathrm{d}}

\newcommand{\figref}[1]{Fig.~\ref{fig:#1}}
\newcommand{\fig}[4]{%
	\begin{figure}[tbp]
		\centering
		\includegraphics[width=#2\textwidth]{#1}
		\caption{\label{fig:#1} #3}
	\end{figure}%
	
}
\newcommand\w[1]{\makebox[2.5em]{$#1$}}
\renewcommand{\sec}[2]{\section{\label{sec:#1}#2}}
\newcommand{\subsec}[2]{\subsection{\label{sec:#1}#2}}
\newcommand{\secref}[1]{Sec.~\ref{sec:#1}}
\newcommand{\eqlabel}[1]{\label{eq:#1}}
\renewcommand{\eqref}[1]{Eq.~(\ref{eq:#1})}

\def\plht{1.5ex}
\newcommand{\plaq}
        {\vcenter{\hbox{\includegraphics[height=\plht]{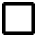}}}}
\newcommand{\dplaqd}
        {\vcenter{\hbox{\includegraphics[height=\plht]{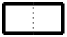}}}}
\newcommand{\dplaqx}
        {\vcenter{\hbox{\includegraphics[height=\plht]{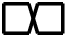}}}}
\newcommand{\dplaql}
        {\vcenter{\hbox{\includegraphics[height=\plht]{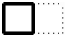}}}}
\newcommand{\dplaqr}
        {\vcenter{\hbox{\includegraphics[height=\plht]{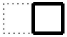}}}}
\newcommand{\dplaqz}
        {\vcenter{\hbox{\includegraphics[height=\plht]{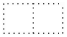}}}}
\newcommand{\SU}{\uparrow}
\newcommand{\SD}{\downarrow}
\newcommand{\SR}{\rightarrow}
\newcommand{\SL}{\leftarrow}

\newcommand{\gen}{\mathcal{S}}
\newcommand{\EE}{S}
\newcommand{\QG}{\mathcal{Q}_8}
\newcommand{\PEQ}{\mathrel{\overset{\makebox[0pt]{\mbox{\normalfont\tiny\sffamily P}}}{=}}}
\newcommand{\Lop}{L}
\newcommand{\Uop}{U}

\begin{document}

\if\draft1 \listofchanges \newpage\fi

\title{Entanglement entropy in lattices with non-abelian gauge groups}

\author{M. Hategan-Marandiuc}
\email{hategan@uchicago.edu}
\affiliation{The University of Chicago, Chicago IL 60637, U.S.A.}

\date{\today}

\begin{abstract}

Entanglement entropy, taken here to be geometric, requires a geometrically 
separable Hilbert space. In lattice gauge theories, it is not immediately clear 
if the physical Hilbert space is geometrically separable. In a previous paper 
we have shown that the physical Hilbert space in pure gauge abelian lattice 
theories exhibits some form of geometric scaling with the lattice volume, which 
suggest that the space is locally factorizable and, therefore, geometrically 
separable. In this paper, we provide strong evidence that indicates that this 
scaling is not present when the group is non-abelian. We do so by looking at 
the scaling of the dimension of the physical Hilbert space of theories with 
certain discrete groups. The lack of an appropriate scaling implies that the 
physical Hilbert space of such a theory does not admit a local factorization. 
We then extend the reasoning, as sensibly possible, to $SU(2)$ and $SU(N)$ to 
reach the same conclusion. Lastly, we show that the addition of matter fields 
to non-abelian lattice gauge theories makes the resulting physical Hilbert 
space locally factorizable.

\end{abstract}

\maketitle
\sec{intro}{Introduction}

There has been considerable debate~\cite{buividovich_entanglement_2008, 
Casini:2013rba, Aoki:2015bsa, Radicevic:2014kqa, Donnelly:2011hn, 
Ghosh:2015iwa, LIN2020115118} about how entanglement entropy can be defined in 
the context of lattice gauge theories. This stems from difficulties in 
reasoning about the physical Hilbert space of such theories when seen through 
the lens of a constrained unphysical space. For example, the unphysical Hilbert 
space of a pure lattice gauge theory is the space of group-valued degrees of 
freedom (\dof) on the links of a lattice. However, constraining the unphysical 
space in order to obtain the physical one leads to a physical Hilbert space of 
lower dimension, and, consequently, fewer \dof. It follows that there would not 
be enough physical \dof\ for every link in the lattice and one must think of 
the physical \dof\ as being associated with different geometrical objects if 
they are to remain on the same lattice. In two spatial dimensions and when the 
gauge group is abelian, the objects with which physical \dof\ are associated 
are the plaquettes. Of course, one does not simply make that choice. Instead, 
one looks at how the physical Hilbert space scales with the lattice size, 
typically using discrete groups, and finds that this scaling depends only on 
the number of plaquettes (see \figref{figures/hs-scaling-links-plaq}). With 
this ansatz, one can derive the precise algebra and form of the physical space. 
A natural question then arises: does this general scheme also apply to 
non-abelian theories? For pure gauge theories, this paper shows that the answer 
is negative. Nonetheless, when matter fields are introduced, the 
non-commutativity of the group elements becomes irrelevant, thus allowing for a 
similar factorization to what is possible with abelian gauge groups when coupled 
with matter.

\sec{overview}{An overview of the problem}

In order to straightforwardly define an entanglement entropy in a quantum 
theory, it is generally necessary for the Hilbert space of the theory to 
exhibit a tensor product structure such that the total space can be expressed 
as a tensor product of local Hilbert spaces~\cite{LIN2020115118, Ghosh:2015iwa, 
Aoki:2015bsa} The local Hilbert spaces form the local \dof\ of the theory. 
Naively, this appears to be the case for lattice theories, in which space is 
discretized and fields are associated with lattice objects, such as vertices, 
links, plaquettes, etc.. In pure gauge Kogut-Susskind Hamiltonian lattice 
theories~\cite{kogutSusskindLattice}, one starts with unphysical gauge fields 
having local unphysical Hilbert spaces associated with links in the lattice. 
Gauge constraints are then imposed on the total unphysical Hilbert space to 
obtain the \emph{physical} Hilbert space as a subspace of the unphysical space. 
It is generally clear that one cannot construct an isomorphism between the 
unphysical and physical spaces on a finite and discrete lattice with discrete 
groups, since the physical space is a strict subspace of the unphysical one. 

What is less obvious is that, when imposing the specific constraints of gauge 
theory, one also cannot generally construct an isomorphism between the set of 
unphysical local Hilbert spaces and the set of physical ones. In other words, 
the physical space will have fewer \dof\ than the unphysical space. This can be 
seen easily if one works with a $Z_2$ gauge group, which results in local 
Hilbert spaces of dimension $2$, the minimally viable dimension of what could 
be considered a \dof. Any reduction in the dimension of the total Hilbert space 
then implies that the number of \dof\ must decrease. The immediate implication 
is that local physical spaces, if they exist at all, cannot generally be 
associated with the same lattice objects as the unphysical ones on a finite 
lattice. A necessary condition for the local physical Hilbert spaces to exist 
is for the dimension of the physical Hilbert space to scale appropriately with 
lattice size. Conversely, a lack of such scaling implies that the total Hilbert 
space cannot be factored into local Hilbert spaces. We will use this to show 
the impossibility of having local physical Hilbert spaces in certain 
non-abelian pure gauge lattice theories.

\fig{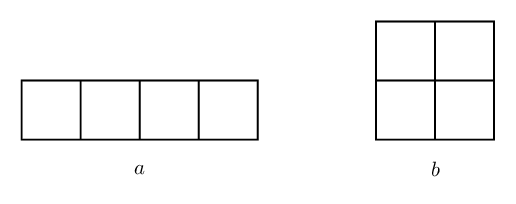}{0.6}{Illustration of the scaling of the 
physical Hilbert space in abelian theories with discrete groups on lattices in 
two spatial dimensions. Both lattices have the same number of plaquettes, but 
lattice (a) has 13 links, whereas lattice (b) has 12 links. It can be shown 
that the physical Hilbert space of both lattices has the same dimension.}

Before moving to the more general lattice problem and entanglement entropy, it 
may be useful to look at some simple systems to illustrate some of the abstract 
points above, as well as to provide a brief overview of some of the more 
significant Hilbert space factorization issues found in literature. We use this 
opportunity to introduce some notation, as well as terminology and basic 
assumptions. We start with a quantum system whose physical space is described 
by a single quantum spin. That is, the physical Hilbert space of this system, 
$\HS^1$, is the space of vectors of the form 
\begin{align}
	\eqlabel{state}
	\ket{\psi} = \alpha \ket{\SU} + \beta \ket{\SD}.
\end{align}
A generating set for the von Neumann algebra of this Hilbert space is
\begin{align}
	\eqlabel{spin-algebra}
	\gen = \{I, \sigma^z, \sigma^x\}.
\end{align}
That is, the von Neumann algebra of the Hilbert space is the smallest von 
Neumann algebra containing $\gen$, and we write this algebra as $\Al(\gen)$. 
The operator $I$ is the identity operator, while $\sigma^z$ and $\sigma^x$ are 
the familiar Pauli operators satisfying $\sigma^z\ket{\SU} = \ket{\SU}, 
\sigma^z\ket{\SD} = -\ket{\SD}$, $\sigma^x\ket{\SU} = \ket{\SD}$, and 
$\sigma^x\ket{\SD} = \ket{\SU}$. However, for consistency with lattice notation 
and clarity, we will relabel the operators as $\Uop \equiv \sigma^x$ and $\Lop 
\equiv \sigma^z$. The operators $\Lop$ and $\Uop$ are reminiscent here of 
canonical variables in the sense that they can be used to completely describe 
the state of the local system at a given time. With that understanding, we will 
call them \emph{canonical pairs}. The algebra generated by canonical pairs, 
together with the identity operator, cannot be factored into a tensor product 
of algebras. We will use the term \emph{local algebra} for such an algebra, 
which is generated by a maximal set of linearly independent non-commuting 
operators, together with the identity, and which cannot be factored. A local 
algebra, together with the local Hilbert space it acts on, is colloquially 
termed \emph{degree of freedom} (\dof). For simplicity, we will consider the 
identity operator as implied when referring to generating sets. An algebraic 
factor, and therefore a local algebra, satisfies $\Al \cap \Al' = \{cI\}$, 
where $\Al'$ is the \emph{commutant} of $\Al$ (the set of all operators that 
commute with all operators in $\Al$) and $c$ is an arbitrary constant. The set 
$\Al \cap \Al' = Z_\Al$ is called the \emph{center} of $\Al$. Algebras that 
satisfy $Z_\Al = \{cI\}$ are said to have a  trivial center and are factors. 
Local algebras, as defined above, are factors by construction.

We can, at this point, consider a larger system, which is formed by taking the 
direct product of two single spin spaces, $\HS^2 = \HS^1_L \otimes \HS^1_R$, 
where the subscripts L and R are used to distinguish the two subspaces. The 
most general state on this space takes the form
\begin{align}
	\ket{\psi}_{LR} = \alpha\ket{\SU_L\SU_R} + \beta\ket{\SU_L\SD_R} 
		+ \delta\ket{\SD_L\SU_R} + \gamma\ket{\SD_L\SD_R},
\end{align}
with the standard normalization condition 
$\alpha^2 + \beta^2 + \delta^2 + \gamma^2 = 1$.

A set of local algebras for this space is $\Al^1_L = \Al(\{\Lop_L \otimes I_R, 
\Uop_L \otimes I_R\}) \equiv \Al(\{\Lop_L, \Uop_L\})$ , $\Al^1_R = 
\Al(\{\Lop_R, \Uop_R\})$, and we omit the multiplication with the identity 
where it can be inferred from the context. This decomposition is only unique up 
to unitary equivalence, and since all Hilbert spaces of the same finite 
dimension are unitarily equivalent, the relevant quantity here is the dimension 
of $\HS^2$. The total algebra of the system $\Al^2$ is the closure of the local 
algebras or the smallest von Neumann algebra that contains both $\Al^1_L$ and 
$\Al^1_R$, and we write $\Al^2 = \Al^1_R \otimes \Al^1_L$. The existence of a 
tensor product factorization of $\HS^2$ or of a factorization of $\Al^2$ do not 
imply that all subalgebras of $\Al^2$ have trivial centers. Conversely, the 
existence of algebras with a non-trivial center does not preclude a 
factorization of the Hilbert space or the algebra. For example, the Hilbert 
space $\HS^2$ can be factorized by construction (i.e., it is a tensor product 
of two Hilbert spaces). However, the algebra 
\begin{align}
	\eqlabel{alz}
	\Al_Z = \Al(\{\Lop_L, \Uop_L, \Lop_R\}),
\end{align}
which is a subalgebra of $\Al^2$, has a non-trivial center since $\Al_Z' = 
\Al(\{\Lop_R\})$ and, therefore, $\Al_Z \cap \Al_Z' = \{c\Lop_R\}$. Even 
simpler, all algebras of the form $\Al(\{A\})$ have a non-trivial center for 
some operator $A$ not proportional to the identity.

The space $\HS^2$ is the same space as the unphysical Hilbert space of a pure 
gauge $Z_2$ $1 + 1$ dimensional Kogut-Susskind Hamiltonian lattice with two 
links. In the gauge theory, one imposes a gauge constraint that restricts the
physical states to states of the form
\begin{align}
	\eqlabel{ent-state}
	\ket{\psi}_{LR}^{\phys} = \alpha\ket{\SU\SU} + \beta\ket{\SD\SD},
\end{align}
where we omitted some of the L and R subscripts. The dimension of the 
resulting physical Hilbert space, which we denote by $\HS^2_{\phys}$, is now 
the same as that of $\HS^1$. The physical Hilbert space has only one \dof\ and 
this \dof\ cannot be meaningfully assigned to the two links in the lattice. The 
gauge constraints act as a cutoff, leading to a change in scale. Imposing the 
constraint that all states take the form in~\eqref{ent-state} means that some 
operators in the initial algebra become unphysical, in the sense that they do 
not preserve this form. For example:
\begin{align}
	\Uop_L (\alpha\ket{\SU\SU} + \beta\ket{\SD\SD}) = 
		\alpha\ket{\SD\SU} + \beta\ket{\SU\SD}.
\end{align}

The operators in $\Al^2$ that preserve the physical form of states are 
generated by the set $S^2_{\phys} = \{\Lop_L, \Lop_R, \Uop_L \Uop_R\}$. The 
operators in $\Al(S^2_{\phys})$ are called \emph{physical operators}. The 
operators $\Lop_L$ and $\Lop_R$ are only distinct when they act on unphysical 
states. For any physical state $\ket{\psi}_{LR}^{\phys}$, 
$\Lop_L\ket{\psi}_{LR}^{\phys} = \Lop_R\ket{\psi}_{LR}^{\phys}$, and we write 
$\Lop_L \PEQ \Lop_R$. To be clear, if we postulate that the physical world 
consists of only physical or gauge invariant states, the operators $\Lop_L$ and 
$\Lop_R$ are the same matrix (i.e., they are the same operator). This 
highlights an important aspect: certain statements about operators in  
$\Al(S^2_{\phys})$ are different when the operators act on the unphysical 
Hilbert space from when they act on the physical subspace. Furthermore, the 
physical algebra cannot resolve more details than can be distinguished through 
the physical Hilbert space. In other words, the subset of operators in $\Al^2$ 
that are physical cannot create or measure more states than are available in 
the physical Hilbert space. That is, there is no L and R in $\HS^2_{\phys}$, no 
more more than there would be in $\HS^1$. We, again, arrive at the idea that 
the dimension of the physical Hilbert space is a significant quantity.

\fig{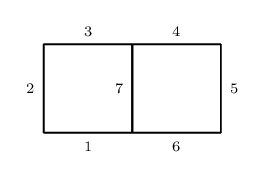}{0.3}{A two plaquette lattice. The numbers label the
links.}

A more pertinent example to lattice gauge theory is the two-plaquette $2+1$ 
dimensional $Z_2$ pure gauge lattice, shown in \figref{figures/two-plaq-lat}. 
The unphysical Hilbert space of this theory associates a $Z_2$ element, which 
is equivalent to a basis of $\HS^1$, to every link in the lattice. The total 
unphysical Hilbert space is a tensor product of $\HS^1$ spaces: $\HS^7 = 
\HS^1_1 \otimes \ldots \otimes \HS^1_7$, where the subscripts label the local 
spaces and correspond to the numbering of the links in 
\figref{figures/two-plaq-lat}. One can, at this point, proceed to (partially) 
fix the gauge using a variation of \emph{maximal tree gauge 
fixing}~\cite{creutz1983quarks} by fixing a set of links to a certain vector 
(e.g., $\ket{\SU}$) as long as the fixed links do not form any loops. This has 
the effect of reducing the dimension of the unphysical Hilbert space, from 
$\dim(\HS^7) = 2^7$ to anything of the form $\dim(\HS) = 2^n,$ with $n \in \{2, 
\ldots, 6\}$, the lower bound being the dimension of the physical space, 
obtained when a maximal set of links has been fixed. The physical space is 
isomorphic to $\HS^2$, which is a well known duality~\cite{wegner}. The physical 
states are of the form
\begin{align}
	\ket{\psi}_{7}^{\phys} = \alpha\ket{\SU\SU\SU\SU\SU\SU\SU}
		+ \beta\ket{\SD\SD\SD\SU\SU\SU\SD}
		+ \gamma\ket{\SU\SU\SU\SD\SD\SD\SU}
		+ \delta\ket{\SU\SU\SU\SU\SU\SU\SD},
\end{align}
and we can identify the unphysical basis as the basis of electric fluxes 
through links. A more illuminating way of writing the physical states is 
\begin{align}
	\ket{\psi}_{7}^{\phys} = \alpha\ket{\dplaqz} + \beta\ket{\dplaql}
		+ \gamma\ket{\dplaqr} + \delta\ket{\dplaqd},
\end{align}
where the thick lines correspond to $\ket{\SD}$ states on the respective 
links. We can see that the space of physical states is isomorphic to an $\HS^2$ 
space with $\ket{\dplaqz} = \ket{\SU\SU}, \ket{\dplaql} = \ket{\SD\SU}, 
\ket{\dplaqr} = \ket{\SU\SD}$, and $\ket{\dplaqd} = \ket{\SD\SD}$. A relevant 
observation is that we cannot associate the two $\HS^1$ factors of $\HS^2$ with 
links in the lattice in any sort of uniform way. Once again, gauge constrains 
act as a cutoff that requires a (slight) change of scale. If one proceeds to 
investigate the physical Hilbert space of larger lattices with the same group 
(and with free boundary conditions), one finds that the physical Hilbert space 
has dimension $\dim(\HS_{\phys}) = 2^n$, where $n$ is the number of plaquettes 
in the lattice. It is then natural to interpret the physical \dof\ as fluxes 
through plaquettes rather than through links.

In terms of algebras, the unphysical algebra can be written as $\Al^7 = 
\Al^1_1 \otimes \ldots \otimes \Al^1_7$, where the subscripts label the links. 
The subset of operators of $\Al^7$ that are physical is generated by 
\begin{align}
	\eqlabel{gset}
	S^7_{\phys} = \{\Lop_i, 
		\Uop_1\Uop_2\Uop_3\Uop_7, 
		\Uop_1\Uop_2\Uop_3\Uop_7\},
\end{align}
where $i \in \{1, \ldots, 7\}$. However, as was the case before, $S^7_{\phys}$ 
contains redundancies such as 
\begin{align}
	\eqlabel{redundancies}
	\Lop_1 &\PEQ\Lop_2 \PEQ \Lop_3, \nonumber\\
	\Lop_4 &\PEQ \Lop_5 \PEQ \Lop_6, \nonumber\\
	\Lop_7 &\PEQ \Lop_3 \Lop_4.
\end{align}

From an algebraic perspective, the question is whether $\Al(S^7_{\phys})$ can 
be factored, and if so, how. The problem is 
typically~\cite{buividovich_entanglement_2008, Casini:2013rba} approached by 
attempting to assign the operators in $S^7_{\phys}$ into two sets, one for each 
of the left and right sides of the lattice. This can naively run into a number 
of issues. Specifically,~\cite{buividovich_entanglement_2008} attempts to 
factor operators such as $\Uop_1\Uop_2\Uop_3\Uop_7$ into $\Uop_1\Uop_2\Uop_3 
I_7$ and $I_1 I_2 I_3\Uop_7$, with $I_n$ being identity operators on the 
respective subspaces, noting that the factored operators cease to act 
exclusively in the physical subspace (i.e., they are not gauge invariant). A 
slightly different attempt is made in~\cite{Casini:2013rba}, which notes 
operator identities similar to those in \eqref{redundancies} and concludes that 
a choice of algebra such as $\Al_{EC} = \Al(\{\Lop_3, \Lop_4, \Lop_7, 
\Uop_4\Uop_5\Uop_6\Uop_7\})$ has $\Lop_4$ as a center, since $\Lop_4$ commutes 
with all other operators in $\Al_{EC}$. This is indeed the case, as it was the 
case with \eqref{alz}. However, this does not imply that all factorization 
choices result in algebras with center. Specifically, $\Al_L = \Al(\{\Lop_3, 
\Uop_1\Uop_2\Uop_3\Uop_7\})$ and $\Al_R = \Al(\{\Lop_4, 
\Uop_4\Uop_5\Uop_6\Uop_7\})$ are both factors and satisfy $\Al(S^7_{\phys}) = 
\Al_L \otimes \Al_R$. Consequently, we can relabel the generating set operators 
as $\Lop_A \equiv \Lop_3, \Uop_L \equiv \Uop_1\Uop_2\Uop_3\Uop_7, \Lop_B \equiv 
\Lop_4,$ and $\Uop_R = \Uop_4\Uop_5\Uop_6\Uop_7$ such that the subalgebras can 
now be written as $\Al_L = \Al(\{\Lop_L, \Uop_L\}), \Al_R = \Al(\{\Lop_R, 
\Uop_R\})$, which matches what one would expect by looking at the dimension of 
the physical Hilbert space.

The change in geometry resulting from the reduction in the dimension of the 
Hilbert space has additional implications to the locality of operators: local 
operators in the physical space are not necessarily local in the unphysical 
space nor are unphysical operators necessarily local in the physical space. For 
example, in the two plaquette lattice, $\Lop_7 \PEQ \Lop_L\Lop_R$ and $\Uop_L = 
\Uop_1\Uop_2\Uop_3\Uop_7$. Both of these equivalences represent matrix 
expressions and, without further constraints, the only meaningful measure of 
locality is the form that a particular matrix takes in a given basis. A more 
meaningful notion of locality arises when dynamics are introduced through a 
Hamiltonian. If the theory being modeled is one that is expected to 
approximate physical reality, then this Hamiltonian should be such that 
correlations do not violate causality, which usually implies that the terms in 
the Hamiltonian are, by construction, local. One may take things a step further 
by noting that it is the dynamics that give the underlying Hilbert space a 
topology. In more specific but perhaps less general terms, the topology and 
geometry of the space are fully given by the way that rays of light propagate 
and the propagation of rays of light is governed by dynamics. The relevance to 
geometric entanglement entropy and Hilbert space tensor product structures is 
that one must be careful when using the locality of operators as a fundamental 
assumption, especially when constraints are involved or when used with theories 
that are not causal.

\subsection{Entanglement Entropy and Unphysical Hilbert Spaces}

Entanglement entropy gives a quantitative measure of the extent to which parts 
of a state belonging to complementary subspaces of a Hilbert space are 
correlated. But physical states in gauge theories, whose Hilbert spaces are 
defined as subspaces of unphysical/extended Hilbert spaces, can appear as 
entangled states in the unphysical space. For example, the state with $\alpha = 
\beta =  1/\sqrt{2}$ in \eqref{ent-state} is a Bell pair with perfect 
entanglement despite the physical space being that of a single spin, for which 
the notion of entanglement seems absurd. There exist 
arguments~\cite{Donnelly:2011hn, buividovich_entanglement_2008} in literature 
that suggest that such entanglement is indeed a legitimate physical phenomenon. 
In particular, for states of the form found in \eqref{ent-state}, one finds that
\begin{align}
	\eqlabel{sa1}
	\EE_1' &= -\tr \rho_1 \log \rho_1 = 
		-\alpha^2 \log \alpha^2 - \beta^2 \log \beta^2,
\end{align}
which is~\cite{Donnelly:2011hn} a classical Shannon entropy term of the 
distribution of basis vectors and where
\begin{align}
	\eqlabel{rhoa1}
	\rho_1 &= \tr_2 \left[
		\alpha^2\ket{\SU}_1\otimes\ket{\SU}_2\bra{\SU}_2\otimes\bra{\SU}_1 + 
		\beta^2\ket{\SD}_1\otimes\ket{\SD}_2\bra{\SD}_2\otimes\bra{\SD}_1 + \cdots
	\right]\nonumber\\
		&= \alpha^2 \ket{\SU}_1\bra{\SU}_1 + \beta^2 \ket{\SD}_1\bra{\SD}_1
\end{align}
is the reduced density matrix.

The need to use unphysical Hilbert spaces to calculate the entanglement 
entropy stems from the assumption that it is impossible to partition the 
physical Hilbert space of lattice gauge theories, an assumption that we have 
shown~\cite{ee-abelian} to be unnecessary in $2+1$ dimensional abelian pure 
gauge theories (briefly illustrated in the previous section) and in abelian 
gauge theories when coupled with matter fields. Furthermore, if we assume that 
the physical Hilbert space is the only accurate representation of physical 
reality, an entanglement entropy calculated on an extended Hilbert space can 
only be valid to the extent that it leads to the same result on any extended 
Hilbert space that starts from the physical space, similar to how the outcome 
of renormalization does not depend on the details of the renormalization 
scheme. However, this is clearly not so with entanglement entropy as we will 
show in the following paragraphs. 

With \eqref{ent-state}, which we repeat here for clarity, we chose to extend 
the physical Hilbert space such that physical states are of the form
\begin{align}
	\eqlabel{ent-state-2}
	\ket{\psi}_{LR}^{\phys} = \alpha\ket{\SU\SU} + \beta\ket{\SD\SD}.
\end{align}
Since there is no physics that can constrain the construction of the 
unphysical space, this is simply a choice of two orthogonal vectors in a 
two-spin space. Consequently, we could have chosen the unphysical space such 
that physical states are of the form
\begin{align}
	\eqlabel{alt-ext-hs}
	\ket{\psi''} = 
		\dfrac{\alpha}{\sqrt{2}}\big(\ket{\SU\SU} + 
		\ket{\SD\SD}\big) + 
		\dfrac{\beta}{\sqrt{2}}\big(\ket{\SU\SU} - 
		\ket{\SD\SD}\big),
\end{align}
for which we would obtain
\begin{align}
	\eqlabel{alt-ext-ee}
	\EE_1'' = 
		-\dfrac{(\alpha + \beta)^2}{2}\log \dfrac{(\alpha + \beta)^2}{2} 
		-\dfrac{(\alpha - \beta)^2}{2}\log \dfrac{(\alpha - \beta)^2}{2}. 
\end{align}
We recognize \eqref{alt-ext-ee} as similar to a momentum-space version of 
\eqref{ent-state-2} and, if the spins in \eqref{ent-state-2} were associated 
with different physical regions of space, there would be a preferential 
geometrical basis that would favor \eqref{ent-state-2} or a local unitary 
transformation of it which would preserve the entanglement entropy. However the 
fact that \eqref{ent-state-2} and \eqref{alt-ext-ee} are related by a unitary 
transformation that acts exclusively in an unphysical subspace, one cannot use 
physical arguments to elevate one choice above the other.

The example physical Hilbert space in \eqref{ent-state-2} can be extended in 
ways that can lead to arbitrarily parameterized entanglement entropies:
\begin{align}
	\ket{\psi''} = \alpha\big(\gamma\ket{\SU\SU}_1\otimes\ket{\SU\SU}_2 + 
		\delta\ket{\SD\SD}_1\otimes\ket{\SD\SD}_2\big) + 
		\beta\big(\gamma\ket{\SU\SD}_1\otimes\ket{\SU\SD}_2 + 
		\delta\ket{\SD\SU}_1\otimes\ket{\SD\SU}_2\big),
\end{align}
with an entanglement entropy of
\begin{align}
	\EE_1'' &= -\alpha^2\gamma^2 \log \alpha^2\gamma^2 -
		\alpha^2\delta^2 \log \alpha^2\delta^2 -
		\beta^2\gamma^2 \log \beta^2\gamma^2 -
		\beta^2\delta^2 \log \beta^2\delta^2 \nonumber\\
		&= -\alpha^2\log\alpha^2 -\beta^2\log\beta^2
			-\gamma^2\log\gamma^2- \delta^2\log\delta^2,
\end{align}
where $\gamma$ can be varied arbitrarily, up to normalization.

Buividovich and Polikarpov~\cite{buividovich_entanglement_2008} argue that a 
``minimal'' extension of the Hilbert space is justified. However, it is unclear 
what ``minimal'' means and whether it can lead to an unambiguous solution. If 
``minimal'' refers to the dimension of the extended Hilbert space, then it does 
not unambiguously specify a solution, as evidenced by \eqref{alt-ext-hs} and 
\eqref{alt-ext-ee}.

These arguments suggest that unphysical Hilbert spaces are unsuitable in 
defining an entanglement entropy, since the results depend on the choice of 
unphysical space and cannot be falsified. This work takes as fundamental the 
assumption that either a definition of entanglement entropy be based 
exclusively on physical aspects of the theory or that the definition is such 
that unphysical aspects can be removed from the final result. Nonetheless, when 
discussing field theories, and, in particular, lattice theories, it is not 
always clear that this is possible, for reasons that we will outline shortly.

\subsec{ee-and-lt}{Geometric Entanglement Entropy and Lattice Theories}

Until this point in the discussion, we focused mostly on a generic form of 
Entanglement Entropy, which applies to any bipartition of the Hilbert space, 
giving little weight to the geometry of the underlying space. However, we are 
often interested in geometric entanglement entropy, which implies a geometric 
bipartition of the Hilbert space in which each subspace belongs to a well 
defined geometric region. In order to be able to talk about subspaces 
associated with arbitrary geometric regions (up to a certain cutoff), the 
dimension of the Hilbert space becomes insufficient. Instead, one must look at 
how this dimension correlates with the volume of such geometric regions.

In field theory, fields are associated with every point in space. We will call 
a \emph{local field theory} a field theory for which homogeneous local algebras 
are associated with every point in bulk space in a translation-invariant way. 
The local algebras can take many forms, and can exhibit additional geometrical 
or internal structure. More generally, one would also allow for algebras 
associated with points on the boundaries of space as well as ``central'' 
algebras, whose operators commute with all other operators in the theory, but 
are not clearly associated with any bulk geometrical objects. The discussion 
can be made significantly more tractable if we enforce two restrictions, of 
which the first is to discretize the space, such as in Hamiltonian lattice 
gauge theories~\cite{kogutSusskindLattice}. This allows us to bypass some of 
the difficulties found in the continuum and treat the problem as a quantum 
mechanical one, with the local algebras acting on local Hilbert spaces, as we 
have done in the previous sections. The second restriction is to focus on local 
Hilbert spaces of finite dimension, which allows us to use simple counting 
arguments. With these assumptions, one expects the total Hilbert space of the 
theory to take the form of a tensor product of the following form:
\begin{align}
	\eqlabel{hs-fact}
	\HS = \bigotimes_{\phi; x \in \mathcal{V}} \HS_{\phi, x}^{\text{bulk}} \otimes
		\bigotimes_{\psi; y \in \mathcal{B}} \HS_{\psi, y}^{\text{boundary}} \otimes
		\bigotimes_{\chi} \HS_{\chi}^{\text{central}},
\end{align}
where $\mathcal{V}$ represents the bulk of the space, $\mathcal{B}$ the 
boundary, $\phi, \psi$, and $\chi$ run over all the possible fields, and 
$\HS_{\phi, \psi, \chi}$ represent local Hilbert spaces on which the respective 
local algebras faithfully act on. For most of what follows, it is also 
sufficient to restrict the discussion to single fields. Then, one can write:
\begin{align}
	\eqlabel{hs-dim-scaling}
	\log \dim(\HS) &= V \log \dim(\HS_{x}^{\text{bulk}}) + 
		B \log \dim(\HS_{y}^{\text{boundary}}) + 
		\log \HS^{\text{central}} \nonumber\\
		&= aV + bB + c,
\end{align}
where $a, b, c \ge 0$ are constants representing the log dimension of the 
local Hilbert spaces and $V$ and $B$ are the volume of the bulk and area of the 
boundary, respectively. The reason why \eqref{hs-dim-scaling} is useful is that 
one does not need to know what the factorization in \eqref{hs-fact} is. One can 
simply look at the scaling of the log-dimension of the Hilbert space with the 
geometrical space. The existence of an appropriate scaling does not necessarily 
guarantee that the algebra of a theory can be factored into a local field 
theory in an obvious or elegant way. On the other hand, if we can count the 
dimension of the physical Hilbert space of a lattice theory and find that it 
does not scale according to \eqref{hs-dim-scaling}, we can conclude that it is 
not a local theory in the sense of \eqref{hs-fact}.

The existence of the factorization in \eqref{hs-fact}, sans the central and 
boundary sectors, enables a textbook definition for entanglement entropy. The 
boundary sector, while not commonly dealt with, is uninteresting if the entire 
geometric space is compact: a subset of the geometric space takes the form $S = 
\mathcal{V}' \cup \mathcal{B}'$, with $\mathcal{V}' \subseteq \mathcal{V}$, 
$\mathcal{B}' \subseteq \mathcal{B}$, and its associated Hilbert space is then 
$\HS_S = \protect\bigotimes_{\phi, x \in \mathcal{V}'}\HS_{\phi, 
x}^{\text{bulk}} \otimes \protect\bigotimes _{\psi, y \in 
\mathcal{B}'}\HS_{\psi, y}^{\text{boundary}}$.. It is less clear, on the other 
hand, what one should do with the central sector and whether one can 
meaningfully talk about entanglement entropy if $c \neq 0$. Surprisingly 
enough, a more interesting situation arises when the log-dimension of the 
Hilbert space of a theory follows the general form of \eqref{hs-fact}, but with 
$c < 0$. This represents a global constraint on the Hilbert space, which yields 
a topological entanglement entropy~\cite{Kitaev:2005dm}. We will come back to 
this point shortly.

\fig{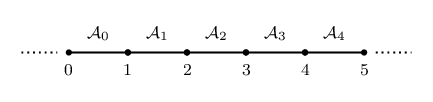}{0.5}{Lattice in one spatial dimension.}

The issue of Hilbert space scaling is largely a trivial issue in theories 
where the physical Hilbert space is postulated, such as scalar theories of 
which the simplest are various quantum spin/Ising models. The more intriguing 
scenario arises in gauge theories, where one starts with an unphysical space 
satisfying the factorization \eqref{hs-fact} with $c = 0$ and $V$ representing 
the number of links in the lattice.  When gauge constraints are imposed, 
however, $\dim(\HS\phys) < \dim(\HS\unphys)$ and the geometric scaling of the 
log-dimension of the physical Hilbert space cannot generally be the same as 
that of the unphysical one. This is immediately apparent if we look at a $1 + 
1$ dimensional lattice gauge theory (see \figref{figures/one-d-lat}) with a 
$Z_2$ group, which shares the algebra seen earlier in \eqref{spin-algebra}. In 
such a lattice, unphysical $Z_2$ local algebras $\Al_i$ are associated with 
every link connecting two nearest vertices. The algebras $\Al_i$ act on local 
unphysical Hilbert spaces, $\HS_i$ whose vectors are denoted by $\ket{u_i}$. 
The \emph{field basis} for each of the local Hilbert spaces consists of vectors 
corresponding to each group element, which we will denote here by $\ket{\SR}$ 
and $\ket{\SL}$, for reasons that will become apparent shortly. A gauge 
transformation would then simultaneously switch basis vectors $\ket{\SR} 
\leftrightarrow \ket{\SL}$ for all links connected to a particular vertex, $i$. 
Ignoring for a moment the remaining links, the most general state invariant 
under such a transformation is:
\begin{align}
	\ket{\psi_i} = 
	\dfrac{\gamma}{\sqrt{2}}\big[
		\ket{\SR}_{i-1}\ket{\SR}_i + 
		\ket{\SL}_{i-1}\ket{\SL}_i\big] +
	\dfrac{\delta}{\sqrt{2}}\big[
		\ket{\SR}_{i-1}\ket{\SL}_i + 
		\ket{\SL}_{i-1}\ket{\SR}_i\big].
\end{align}
If we make a local basis change defined by $\ket{\SU} = \dfrac{1}{\sqrt{2}} 
\left(\ket{\SL} + \ket{\SR}\right)$ and $\ket{\SD} = \dfrac{1}{\sqrt{2}} 
\left(\ket{\SL} - \ket{\SR}\right)$, we obtain:
\begin{align}
	\ket{\psi_i} = \dfrac{\gamma + \delta}{\sqrt{2}}\ket{\SU}_{i-1}\ket{\SU}_i
		+ \dfrac{\gamma - \delta}{\sqrt{2}}\ket{\SD}_{i-1}\ket{\SD}_i.
\end{align}
But this is precisely the form of the state in \eqref{ent-state} with $\gamma 
= (\alpha + \beta) / \sqrt{2}$ and $\delta=(\alpha - \beta) / \sqrt{2}$, which 
we restate:
\begin{align}
	\ket{\psi'} = \alpha \ket{\SU}_{i-1}\ket{\SU}_i + 
		\beta \ket{\SD}_{i-1}\ket{\SD}_i.
\end{align}
This equation is Gauss' law (for the $Z_2$ group): in the absence of charges, 
whatever we measure for the electric field on link $i-1$ will also be measured 
on link $i$. If we further require gauge invariance at every vertex in the 
lattice, we arrive at a general form for physical states on a $1+1$ dimensional 
$Z_2$ lattice:
\begin{align}
	\ket{\psi}_{\text{phys.}} = 
		\alpha \big(\cdots \otimes\ket{\SU}_{0}\otimes\ket{\SU}_1\otimes \cdots 
			\otimes \ket{\SU}_5 \otimes \cdots\big) + 
		\beta \big(\cdots \ket{\SD}_{0}\otimes\ket{\SD}_1\otimes \cdots 
			\otimes \ket{\SD}_5 \otimes\cdots\big),
\end{align}
or, more compactly,
\begin{align}
	\eqlabel{phys-state-1dz2}
	\ket{\psi}_{\text{phys.}} = 
		\alpha \ket{\cdots \SU\SU \cdots \SU\cdots} + 
		\beta \ket{\cdots \SD\SD \cdots \SD\cdots}. 
\end{align}

As expected, we end up with a physical space that is of a lower dimension than 
the unphysical space. Whereas the unphysical space satisfied $\log \dim \HSu = 
V\log 2$, the physical one follows $\log \dim \HSp = \log 2$. This is 
significant because it implies that we cannot associate a physical $Z_2$ 
algebra with every link in the lattice, nor can we split the physical Hilbert 
space in a way that would allow us to associate portions of it with every link. 
It should be noted that, since the physical algebra cannot be used to create or 
measure states that are outside of the physical Hilbert space, there is no 
intrinsic metric that would allow us to distinguish a finer geometry than what 
the dimension of the physical Hilbert space allows. In other words, one cannot 
meaningfully speak of geometric entanglement entropy below the distance cutoff 
of the theory, which, in this case, is the whole lattice.

There is little to be learned from the one dimensional case for lattices in 
higher dimensional pure gauge theories. In two or more dimensions, bulk 
vertices connect more than two links, making Gauss' law more complex. In two 
dimensions, the physical Hilbert space can be described in terms of closed 
electric curves, but the precise details depend on the boundary 
conditions~\cite{ee-abelian}. With free boundary conditions, the dimension of 
the physical Hilbert space scales exactly with the area of the lattice 
expressed as the number of plaquettes. Since all states consist of 
superpositions of closed electric loops, one can more readily describe the 
physical space as the space of magnetic fluxes going through plaquettes. This 
is the simple case. When the boundary conditions are such that the lattice 
acquires the topology of a closed surface, one encounters the magnetic Gauss' 
law, which imposes a global constraint on the now unphysical Hilbert space of 
magnetic fluxes through plaquettes. For example, consider the Hilbert space of 
magnetic fluxes through tiles of a sphere for some tiling of the sphere. A 
vector in this basis is fully described by the fluxes through all but one tile, 
a tile which can be chosen at random. The flux through this tile can be 
obtained through the magnetic Gauss' law. However, with a tile removed, the 
spherical symmetry (or a polyhedral approximation of it) is lost. It is not 
immediately clear if this Hilbert space has a spherically symmetric formulation 
which would allow it to be interpreted as a field theory in which independent 
local algebras are associated with every unit of area or tile. If one now 
considers a state such as
\begin{align}
	\eqlabel{low-coupling-vacuum}
	\ket{\psi}_{c} = K \sum_{m_i \in \mathcal{B}_{\text{mag.}}(\HS)} \ket{m_i},
\end{align}
where $K$ is a normalization constant and $\mathcal{B}_{\text{mag.}}(\HS)$ is 
the basis of magnetic fluxes through plaquettes such that all $m_i \in 
\mathcal{B}_{\text{mag.}}(\HS)$ satisfy Gauss' law, one obtains a constant 
entanglement entropy that does not depend on the size or shape of the 
entanglement region. The resulting entanglement entropy is 
called~\cite{Kitaev:2005dm} \emph{topological entanglement entropy}. The state 
in \eqref{low-coupling-vacuum} is often called \emph{the low coupling ground 
state}, for reasons that we will not discuss here.

It is notable that, in the sphere example, topological entanglement entropy is 
also independent of scale, provided that the state remains one in which there 
is no entanglement entropy when calculated on the unconstrained space (i.e., 
the low coupling ground state). One can use progressively finer tilings and 
obtain the same result. It can then be asked whether this property is preserved 
when the continuum limit is taken. In the continuum limit, the tile in A would 
have a vanishing area. In~\cite{Callan:1994py}, it is argued that the field at 
a single point would have zero integration measure and be irrelevant to the 
calculation. Naively extending that reasoning would suggest that the continuum 
limit of the entanglement entropy of the low coupling ground state of the 
magnetic theory on a sphere should be zero, in apparent conflict with the scale 
independence of the topological entanglement entropy. A similar 
observation can be found in~\cite{delcamp16onee}.

There is an important conclusion that can be drawn at this point which is 
that, despite certain difficulties with the log-dimension scaling of the 
physical Hilbert space related to non-zero coefficients for boundary and 
central algebras in \eqref{hs-dim-scaling}, abelian gauge theories exhibit a 
linear scaling with bulk volume, which suggests that, in many cases, their 
physical algebra can be factored as a local algebra. This appears to be 
fundamentally different from non-abelian pure gauge theories. In the next 
section, we will provide evidence that non-abelian theories have a complex bulk 
scaling which prevents such a factorization. We have shown in~\cite{ee-abelian} 
that coupling to matter fields elegantly solves the factorization problem in 
abelian theories. It does so regardless of dimension and boundary conditions: 
given any graph (in the graph theoretical sense) with electric fluxes on 
edges/links, one can always find the necessary charges at vertices to satisfy 
Gauss' law. In \secref{na-matter}, we will show that, unlike with pure 
non-abelian gauge fields, the factorization carries over to non-abelian gauge 
theories when coupled to matter fields.

\section{Pure non-abelian gauge theories}

The main goal of this section is to investigate the scaling of the 
log-dimension of the physical Hilbert space with lattice size in pure 
non-abelian lattice gauge theories. The Hilbert space dimension is, 
unfortunately, only well defined for discrete groups. However, we provide some 
evidence that suggests that the primary conclusion that can be derived from 
discrete groups, the non-existence of a local factorization of the algebra of 
the theory, likely extends to continuous groups. We discuss how this lack of 
factorization relates to the familiar ways of looking at the Hilbert space of 
nonabelian lattice gauge theories.

The counting of degrees of freedom is complicated by the existence of local 
gauge transformations. In order to make the problem tractable, we will use 
maximal tree gauge fixing~\cite{creutz1983quarks}, which, in nonabelian 
theories, reduces the degeneracy to a global gauge symmetry.

\subsection{Preliminaries}

We start by introducing the basic formalism that we will use when working with 
Hamiltonian lattice theories. For the derivation of Hamiltonian lattice theory 
from Wilson's lattice theory, see, for example,~\cite{kogutSusskindLattice} 
and~\cite{creutzHamiltonianLattice}. The vertices in the lattice are denoted as 
points $x$ with coordinates
\begin{align}
	x_i = a n_i, a \in \mathbb{R}, n_i \in \mathbb{N},
\end{align}
where $a$ represents the unit spacing of the lattice, and $i$ labels spatial 
dimensions. We will generally use lattice units $a = 1$ and not explicitly 
mention $a$.

Matter fields are associated with vertices, and are denoted by a standard 
Greek letter and a vertex, such as $\phi(x)$. The local unphysical Hilbert 
space of the matter fields $\HS_\phi(x)$ is the space of square integrable 
functions on $\phi(x)$ and vectors in the possibly non-normalizable field basis 
are denoted by, e.g., $\ket{\phi(x)}$.

Gauge fields are associated with links connecting nearest-neighboring 
vertices. They are denoted by $u_i(x) \in G$, where $x$ represents a vertex at 
one end of the link, $i$ is the spatial direction in which the other end of the 
link is to be found, and $G$ is the gauge group. That is, the field $u_i(x)$ 
connects the vertices at $x$ and $x + \hat{i}$. The local unphysical Hilbert 
space of the gauge field $\HS_{u_i(x)}$ is the space of square integrable 
functions of unit norm on $G$ and vectors in the field basis are denoted by 
$\ket{u_i(x)}$.

We define the operators $\Phi(x)$ and $U_i(x)$ that are diagonal in the field 
basis of the matter and gauge fields, respectively:
\begin{align}
	\Phi(x)\ket{\phi(x)} &= \phi(x)\ket{\phi(x)} \\
	U_i(x)\ket{u_i(x)} &= u_i(x)\ket{u_i(x)}\\
	U_{-i}(x + \hat{i})\ket{u_i(x)} &= u^\dagger_i(x)\ket{u_i(x)},
\end{align}
The values $\phi(x)$ and $u_i(x)$ outside the kets are to be understood as 
abstract objects whose multiplication with the kets is not always well defined. 
They can be used to construct states only when appropriate wavefunctions from 
their domain to the complex numbers are specified. The total unphysical Hilbert 
space is obtained as a tensor product of all local Hilbert spaces:
\begin{align}
	\eqlabel{hs-lat-th}
	\HS = \HS_\phi \otimes \HS_u = \left(\bigotimes_x \HS_{\phi(x)}\right)
		\otimes \left(\bigotimes_{x', i} \HS_{u_i(x')}\right),
\end{align}
where $x'$ and $i$ are such that links do not extend past the boundary of the 
lattice, if such a boundary exists.

Gauge transformations are families of operators $\GT[g(x)], g(x) \in G$ that 
associate a group element with each vertex, where we used the square brackets 
to indicate that $\GT[]$ depends on the $g(x)$ for all $x$. The fields transform 
as follows under gauge transformations:
\begin{align}
	\eqlabel{gt-matter-sector}
	\GT[g(x)]\ket{\phi(x)} &= \ket{g(x) \phi(x)} \\
	\eqlabel{gt-gauge-sector}
	\GT[g(x)]\ket{u_i(x)} &= \ket{g(x) u_i(x) g^{-1}(x + \hat{i})}.
\end{align}

A \emph{local gauge transformation} is a gauge transformation $\GT[I, I, 
\ldots, g(x_0), \ldots, I]$ for which all except one $g(x_0)$ are set to the 
group identity. Local gauge transformations are, therefore, associated with a 
single vertex.

The physical Hilbert space of the theory is the space of vectors invariant 
under \emph{all} gauge transformations:
\newcommand{\HSP}{\HS^{\text{phys.}}}
\begin{align}
	\HSP = \big\{ \ket{\psi} = \ket{\psi}_\phi \otimes \ket{\psi}_u \in \HS \,\big\vert\, 
		\GT[g(x)]\ket{\psi} = \ket{\psi}, \forall g(x) \in G \big\}.
\end{align}

The \emph{gauge orbit} of $\ket{\psi}$ is the set of vectors obtained by 
applying all possible gauge transformations to it:
\begin{align}
	O[\psi] = \big\{\ket{\psi'} \in \HS \big| \ket{\psi'} = \GT[g(x)]\ket{\psi}, 
		\forall g(x) \in G\big\}.
\end{align}
The space of gauge orbits and $\HSP$ are isomorphic. Consequently, one can 
find the dimension of the physical Hilbert space by finding a maximal set of 
vectors in the field basis of the unphysical space such that no two vectors in 
the set can be related by a gauge transformation.

In pure gauge theories, there is no matter field and $\HS = \HS_u$ in 
\eqref{hs-lat-th}, with gauge transformations being described exclusively by 
\eqref{gt-gauge-sector}.

A useful class of operators is that of Wilson loops, which are constructed from 
products of operators $U_i(x)$ taken around closed loops:
\begin{align}
	\eqlabel{wilson-loop}
	U_\MC 
		&= U_{i_1}(x_1)U_{i_2}(x_2) \cdots U_{i_n}(x_n),
\end{align}
where
\begin{align}
	\eqlabel{wl-ce1}
	x_{k + 1} &= x_k + \hat{i_k} \\
	\eqlabel{wl-ce2}
	x_1 &= x_n + \hat{i_n},
\end{align}
with $x_k$ being vertices along the closed curve $\MC$. By applying 
$U_\MC$ to gauge transformed fields from \eqref{gt-gauge-sector}, one 
obtains its transformation properties under a gauge transformation:
\begin{align}
	U_\MC \bigotimes_{k, y}\ket{u'_{k}(y)} &=
		U_{i_1}(x_1)U_{i_2}(x_2) \cdots U_{i_n}(x_n) \bigotimes_{k, y}
				\ket{g(y) u_k(y) g^{-1}(y + \hat{k})} \nonumber\\
				&= g(x_1) u_{i_1}(x_1) g^{-1}(x_1 + \hat{i}_1) 
					g(x_2) u_{i_2}(x_2) g^{-1}(x_2 + \hat{i}_2) \cdots \nonumber\\
				&\quad\quad\quad g(x_n) u_{i_n}(x_n) g^{-1}(x_n + \hat{i}_n) 
					\bigotimes_{k, y}\ket{g(y) u_k(y) g^{-1}(y + \hat{k})}.
\end{align}
Applying \eqref{wl-ce1} and \eqref{wl-ce2}, we obtain
\begin{align}
	U_\MC \bigotimes_{k, y}\ket{u'_{k}(y)} = g(x_1) u_{i_1}(x_1) u_{i_2}(x_2) \cdots 
					u_{i_n}(x_n) g^{-1}(x_1) \bigotimes_{k, y}\ket{u'_{k}(y)}.
\end{align}
We can then take any function $f_c : G \rightarrow \mathbb{C}$ that satisfies 
$f_c(gug^{-1}) = f_c(u), \forall g, u \in G$ (i.e., a class function on $G$) 
and verify that
\begin{align}
	\GT^{-1} f_c(U_\MC) \GT\bigotimes_{k, y}\ket{u_{k}(y)} 
		&= f_c(U_\MC) \bigotimes_{k, y}\ket{u_{k}(y)},
\end{align}
hence $f_c(U_\MC)$ is a gauge invariant operator. The most common 
$f_c$ is the trace of a group element in some representation of the group and 
the operator $\tr(U_\MC)$ is a \emph{Wilson loop}.

\subsection{Electric states}

The physical space of gauge theories is often described in terms of states of 
definite electric flux constrained by Gauss' law. It can be shown that these 
states must take the form
\begin{align}
	\ket{\psi^l} = \dfrac{1}{\sqrt{|G|}}\sum_{g \in G}\lambda^l(g)\ket{g},
\end{align}
where $\lambda^l$ is a one-dimensional representation of the group labeled by 
the electric flux $l$ and $|G|$ is the dimension of the group. For continuous 
groups, the sum is replaced by a Haar integral. Given two links with a shared 
vertex (i.e., a one-dimensional theory), gauge invariance at that vertex 
implies that
\begin{align}
	\GT[h] \ket{\psi^l} \otimes \ket{\psi^m} &= 
		\GT[h] \dfrac{1}{|G|}\sum_{g_1, g_2 \in G} \lambda^l(g_1) \lambda^m(g_2) \ket{g_1} \otimes \ket{g_2} \nonumber\\
		&= \dfrac{1}{|G|}\sum_{g_1, g_2 \in G} \lambda^l(g_1) \lambda^m(g_2) \ket{g_1 h^{-1}} \otimes \ket{h g_2} \nonumber\\
		&= \dfrac{1}{|G|}\sum_{g_1', g_2' \in G} \lambda^l(g_1')\lambda^l(h)\lambda^m(h^{-1}) \lambda^m(g_2') \ket{g_1'} \otimes \ket{g_2'} \nonumber\\
		&= \dfrac{1}{|G|}\sum_{g_1', g_2' \in G} \lambda^l(g_1')\lambda^m(g_2') \ket{g_1'} \otimes \ket{g_2'}.
\end{align}
This can only be true if $\lambda^l(h)\lambda^m(h^{-1}) = 1$ for all $h$, 
which implies $l = m$. In other words, physical states are states in which the 
electric flux must be conserved. This is Gauss' law. For finite abelian groups, 
the number of orthogonal one-dimensional representations equals the dimension 
of the group. Consequently, electric states on links form a basis for the 
unphysical space. The physical space can then obtained by imposing Gauss' law 
on the electric states. For non-abelian groups, the number of one-dimensional 
representations is equal to the number of conjugacy classes, which is less than 
the dimension of the group. This makes it impossible to use constrained 
electric states as a basis for physical states in non-abelian theories. We will 
give a slightly more detailed version of this statement when talking about the 
Quaternion group.

\subsection{Maximal Tree Gauge Fixing}

\fig{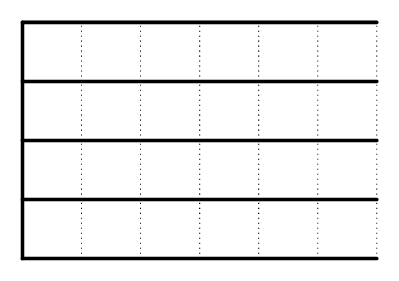}{0.3}{A maximal tree on a 2-d lattice. Thick links are set 
to the identity. In two dimensions with free boundary conditions, there are as 
many remaining links as there are plaquettes.}

Maximal tree gauge fixing (see~\cite{creutz1983quarks}) involves fixing a set 
of links in the lattice to a specific vector, which is most conveniently taken 
to be $\ket{I}$, where $I \in G$ is the identity element of the group. The set 
of links is such that they form a maximal tree, which, by definition, is a set 
of links to which the addition of any other link would result in the creation 
of a loop. An example of a maximal tree is shown in \figref{figures/mtgf}. For 
nonabelian gauge theories, this type of gauge fixing leaves an ancillary global 
gauge transformation, $\GT[g, g, \ldots, g]$ (or $\GT[g]$, in short). Under 
such a gauge transformation, link vectors in the field basis transform as
\begin{align}
	\eqlabel{gt}
	\ket{u_i(x)} \rightarrow \ket{g u_i(x) g^{-1}}.
\end{align}
It is clear that links for which $u_i(x) = g_z, g_z \in Z(G)$, where $Z(G)$ is 
the center of the group, remain invariant under such transformations. When $g$ 
spans $G$ in \eqref{gt}, the remaining links span the respective conjugacy 
classes of the group. Naively, one might be led to the conclusion that the 
physical Hilbert space is isomorphic to the space obtained from the tensor 
product of the local spaces of the group conjugacy classes $\Cl(G)$ on every 
unfixed link. However, as will be shown, this is not generally true.

When considering electric states on maximal tree gauge fixed lattices, each 
unfixed link can be treated independently and there is no Gauss' law to enforce 
in either abelian or non abelian theories. This is because a global gauge 
transformation acts on electric states as follows:
\begin{align}
	\GT[h] \dfrac{1}{\sqrt{|G|}}\sum_{g \in G} \lambda^l(g) \ket{g} 
		&= \dfrac{1}{\sqrt{|G|}}\sum_{g \in G} \lambda^l(g) \ket{h g h^{-1}} \nonumber\\
		&= \dfrac{1}{\sqrt{|G|}}\sum_{g' \in G} \lambda^l(h^{-1} g' h) \ket{g'} \nonumber\\
		&= \dfrac{1}{\sqrt{|G|}}\sum_{g' \in G} \lambda^l(g') \ket{g'},
\end{align}
where we used the fact that $\lambda^l$ is a one-dimensional representation to 
commute $g'$ and $h$. Consequently, electric states are automatically invariant 
under global gauge transformations. This implies that if electric states form a 
basis for local physical states, one could express the total Hilbert space as a 
tensor product of electric states on unfixed links in a maximal tree gauge 
fixed lattice.

\subsection{The Quaternion Group}

\newcommand{\mset}[1]{\{#1\}}

The quaternion group ($\QG$) is one of the simplest nonabelian 
groups. One way to represent it is with the familiar Pauli matrices:
\begin{align}
	\QG = \{\pm I, \pm i\sigma_a\}, a \in \mset{1, 2, 3},
\end{align}
with $I$ being the $2 \times 2$ identity matrix. It has five conjugacy classes:
\begin{align}
	\Cl(\QG) = \big\{\mset{+I}, \mset{-I}, \mset{+i\sigma_a, -i\sigma_a}\big\}, 
		a \in \mset{1, 2, 3}.
\end{align}

\fig{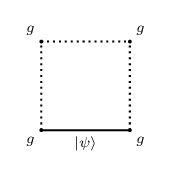}{0.25}{Illustration of physical states on a non 
abelian single plaquette lattice with maximal tree gauge fixing. Dotted links 
are fixed to the identity group element. A global gauge transformation by some 
element $g$ at all vertices is still possible.}

One can construct an algebra diagonal in the conjugacy classes of the group by 
following the general idea behind Wilson loops. In the above representation, a 
set of operators generating this algebra is:
\begin{align}
	S_{\Al_U} = \big\{
		U_\MC^0 \equiv \frac{1}{2}\tr U_\MC, 
		U_\MC^a \equiv \frac{1}{4}\tr^2(i\sigma_a U_\MC)\big\},  
		a \in \mset{1, 2, 3},
\end{align}
where $\MC$ represents ordered links forming a closed curve. On a one-plaquette 
lattice, the dimension of the physical Hilbert space is five, corresponding to 
the number of conjugacy classes of the group:
\begin{align}
	\HS\phys_{\plaq} = \lspan \mathcal{B}_{\plaq, U}\phys = \lspan \big\{\ket{+I}, \ket{-I}, 
		\frac{1}{\sqrt{2}}(\ket{+i\sigma_a} + \ket{-i\sigma_a})\big\}, a \in \{1, 2, 3\},
\end{align}
where the square symbol denotes a quantity belonging to a single plaquette. 
This is illustrated in \figref{figures/onep-phys-hs}, which fixes the gauge by 
setting three of the four plaquette links to $\ket{I}$. A remaining global 
gauge transformation by some group element $g \in \QG$ leaves the identity 
links ($1, 2,$ and $3$) unchanged, while rotating the remaining link through 
its conjugacy class. However, if the state of the unfixed link is in 
$\HS\phys_{\plaq}$, all links remain invariant under all gauge transformations.

The action of $S_{\Al_{\plaq}}$ on states of the form $\ket{g}, g \in \QG$ 
attached to the unfixed link of a plaquette (e.g., 
\figref{figures/onep-phys-hs}), with the square symbol denoting the 
conter-clockwise curve formed by the links of the plaquette, is:

\begin{align}
	U_{\plaq}^0 \ket{\pm I} &= \pm\ket{\pm I}, \nonumber\\
	U_{\plaq}^0 \ket{\pm i\sigma_a} &= 0, a \in \mset{1, 2, 3}, \nonumber\\
	U_{\plaq}^b \ket{\pm I} &= 0, \nonumber\\
	U_{\plaq}^b \ket{\pm i\sigma_a} &= \delta_{ab} \ket{\pm i \sigma_a}, a \in \mset{1, 2, 3}.
\end{align}

\fig{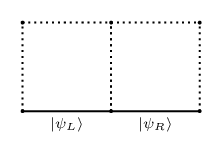}{0.3}{Illustration of physical states on a non 
abelian two-plaquette lattice with maximal tree gauge fixing. Dotted links are 
fixed to the identity group element.}

On a two-plaquette lattice (see \figref{figures/twop-phys-hs}), maximal tree 
gauge fixing leads to two free links. One would, therefore, expect that $\dim 
\HS\phys = \left(\HS\phys_{\plaq}\right)^2 = 25$. A careful count of the gauge 
orbits, however, reveals that there are three additional states. These states 
arise from the fact that the single plaquette orbits are not necessarily 
separable, resulting in the following states being distinct physical states:
\def\qgket#1#2{\ket{#1 i\sigma_#2}}
\def\qgdket#1#2#3#4{\ket{#1 i\sigma_#2}\ket{#3 i\sigma_#4}}
\begin{align}
	\eqlabel{qg-extended-state-functions}
	\ket{\psi_{1, a}} &= \dfrac{1}{\sqrt{2}}\left(
		\qgket+a_{(L)}\qgket+a_{(R)} + \qgket-a_{(L)}\qgket-a_{(R)}\right), \nonumber\\
	\ket{\psi_{2, a}} &= \dfrac{1}{\sqrt{2}}\left(
		\qgket+a_{(L)}\qgket-a_{(R)} + \qgket-a_{(L)}\qgket+a_{(R)}\right),
\end{align}
and we omitted the range of $a$, which is as before. One can see that the two 
are distinct states by acting on them with $U_{\dplaqd}^0$, which is $U_\MC^0$ 
with $\MC$ going around both plaquettes:
\begin{align}
	\eqlabel{qg-extended-states}
	U_{\dplaqd}^0 \ket{\psi_{1, a}} 
		&= \dfrac{1}{\sqrt{2}}\left[\frac{1}{2}\tr (-\sigma_a^2) \qgdket+a+a + 
			\frac{1}{2}\tr(-\sigma_a^2) \qgdket-a-a\right] \nonumber\\
		&= -\ket{\psi_{1, a}}, \nonumber\\
	U_{\dplaqd}^0 \ket{\psi_{2, a}} 
		&= \dfrac{1}{\sqrt{2}}\left[\frac{1}{2}\tr (\sigma_a^2) \qgdket+a-a + 
			\frac{1}{2}\tr(\sigma_a^2) \qgdket-a+a\right] \nonumber\\
		&= +\ket{\psi_{2, a}}.
\end{align}
As can be easily verified, these states are indistinguishable using local 
physical operators $U_{\dplaql}^0, U_{\dplaqr}^0, U_{\dplaql}^a, U_{\dplaqr}^a$ 
which go around either of the plaquettes. Alternatively, and with some matrix 
arithmetic which we will not reproduce here, one can see that the two families 
of states do not mix under the ancillary global gauge transformation. It is 
quite clear that the $28$ physical states in the two plaquette lattice cannot 
belong to a homogeneous tensor product (i.e., a tensor product in which all 
factors have the same dimension). Even if the physical degrees of freedom were 
not associated with plaquettes, it remains clear that the number $28$ is not an 
integral power of any integer and cannot represent the dimension of a tensor 
product of homogeneous local Hilbert spaces.

The physical Hilbert space on the one plaquette lattice, as seen above, can be 
described in terms of the conjugacy classes of the group. We can separate the 
classes into the ``poles'', $P = \lspan\big\{\ket{I}, \ket{-I}\big\}$ and the 
``bulk``, $B = \lspan\big\{\frac{1}{\sqrt{2}}(\ket{+i\sigma_a} + 
\ket{-i\sigma_a}\big\}, a \in \{1, 2, 3\}$. Then, $\HS\phys_{\plaq} = P \oplus 
B$. On two plaquettes, we can classify the states based on the action of the 
Wilson loop operators on the plaquettes as well as around both of the 
plaquettes. With a factorizable space, we would expect $\HS\phys_{\dplaqd} = (P 
\otimes P) \oplus (B \otimes P) \oplus (P \otimes B) \oplus (B \otimes B)$. 
However, we have seen that some of the states in $B \otimes B$ acquire a 
splitting and the last term becomes isomorphic\footnote{The precise form of the 
term is less important, since we care mostly about counting states.} to $(B 
\otimes B) \oplus X$, for some $X$. Using counting arguments alone, it would 
still be possible to factorize such an enlarged space. However, since the two 
plaquette space is larger than the product of single plaquette spaces, 
homogeneous factors would also have to be larger than $\HS\phys_{\plaq}$. The 
smallest such factors would be of dimension $\dim\HS\phys_{\plaq} + 1$, which 
implies that $\dim X > 2 \dim\HS\phys_{\plaq}$. This is not the case for the 
quaternion group, since $\dim X = 3 \ngtr 10 = 2\dim\HS\phys_{\plaq}$.

Numerical calculations\footnote{The code is available at 
\url{https://github.com/hategan/phys-hs-scaling-na}} of the Hilbert space 
dimension are shown in Table~\ref{tab:qg-hs-scaling} and, graphically, in 
\figref{plots/dof-scaling-q8}. One can infer that the dimension of the physical 
Hilbert space\footnote{This scaling holds for the data shown and does not 
necessarily extend to larger lattices.} takes the form $\dim \HS\phys = 4^{n_p 
- 1} \times (2^{n_p} + 3)$ hence
\begin{align}
	\log\dim\HS\phys = (n_p - 1) \log 4 + \log(2^{n_p} + 3).
\end{align}
This scaling cannot be fit in the context of what one would expect from a 
theory with a geometric tensor product structure, as suggested by 
\eqref{hs-dim-scaling}, which requires a linear scaling of the log-dimension of 
$\HS\phys$ with lattice size.

\begin{table}[htb]
\setlength{\tabcolsep}{12pt}
\begin{filecontents*}{plots/dof-scaling-q8.csv}
1,5,5,8
2,28,25,64
3,176,125,512
4,1216,625,4096
5,8960,3125,32768
6,68608,15625,262144
7,536576,78125,2097152
8,4243456,390625,16777216
\end{filecontents*}
\csvreader[
	tabular =  c r r r,
	table head = \# of plaquettes. & $\dim \HS\phys$ & $\dim \HS^{\text{abel.}} $ & $\dim \HS\unphys$\\\hline,
	late after line = \\,
	head = false
]{plots/dof-scaling-q8.csv}{}{%
	\csvcoli & \csvcolii & \csvcoliii & \csvcoliv
}

\caption{This table shows the scaling of the physical Hilbert space 
($\HS\phys$) of a two-dimensional lattice gauge theory with a quaternion group. 
For comparison purposes, also shown are the scaling of the dimension of the 
unphysical Hilbert space ($\HS\unphys$) on a maximal tree gauge fixed lattice 
and the scaling of the dimension of $\HS^{\text{abel.}}$, a physical Hilbert 
space with an abelian group having a dimension equal to the dimension of the 
single-plaquette quaternion physical space (i.e., $Z_5$).}

\label{tab:qg-hs-scaling}
\end{table}

\fig{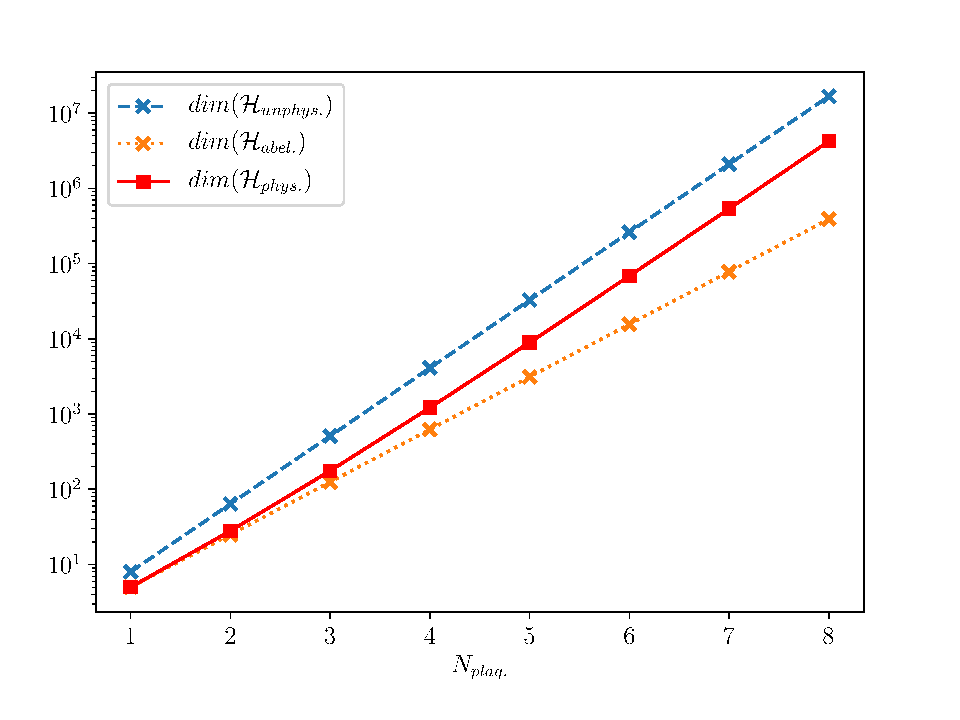}{0.9}{A plot of the data in 
Table~\ref{tab:qg-hs-scaling}. As the number of plaquettes increases, the 
non-local physical states become dominant.}

\fig{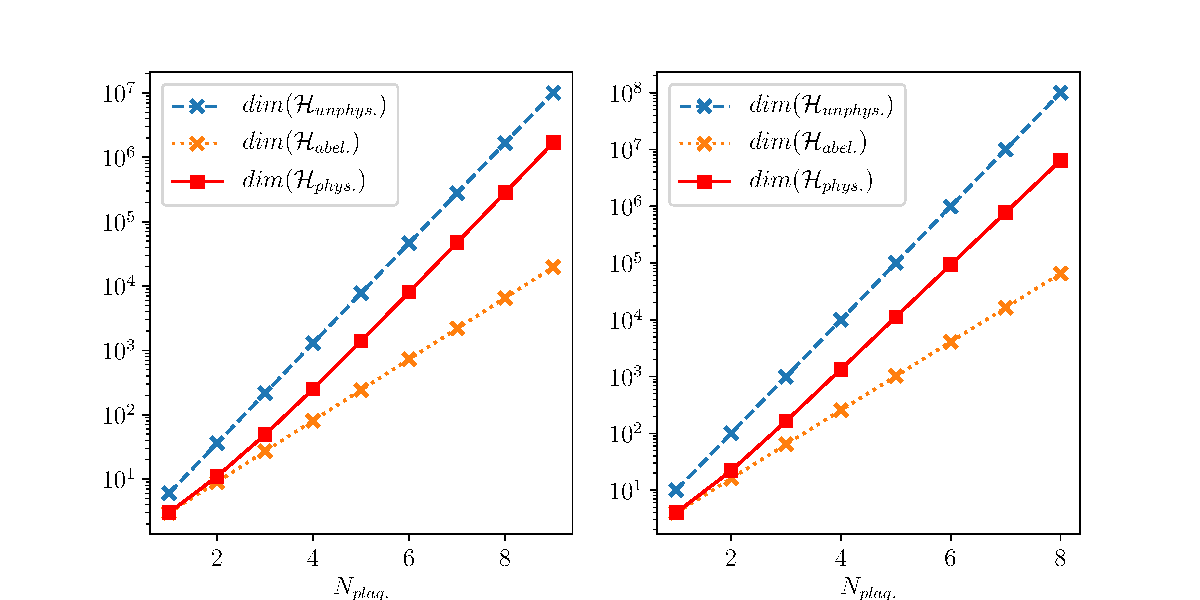}{0.9}{Scaling of the physical Hilbert space 
in a two-dimensional lattice gauge theory with the dihedral groups $D_3$ (left) 
and $D_5$ (right).}

An identical scaling as with the quaternion group is obtained with the 
dihedral group $D_4$, which shares the order and number of conjugacy classes 
with $\QG$. Similar results are obtained with the dihedral groups $D_3$ and 
$D_5$, and the scaling of the corresponding Hilbert spaces can be seen in
\figref{plots/dof-scaling-d-combined}.

It can be noted that a number of solutions proposed in literature that attempt 
to address the geometric non-separability of the physical Hilbert space of 
lattice gauge theories do not address the existence of the non-local states 
seen here. Gauge fixing by setting links at the boundary of the entanglement 
region to the group identity, as suggested in~\cite{Casini:2013rba}, does not 
change any of the arguments above if the boundary of the entanglement region is 
part of the maximal tree used in gauge fixing. The non-local states are not 
restricted to neighbouring plaquettes, and this issue is not addressed by  
various Hilbert space extensions schemes at the 
boundary~\cite{buividovich_entanglement_2008, Radicevic:2014kqa}. To see this, 
one can consider two arbitrarily space-separated gauge links, in a lattice with 
maximal tree gauge fixing and states in which all remaining links except the 
two are set to the group identity, as in \figref{figures/big-lattice-states}. 
Then, one can consider the states in \eqref{qg-extended-state-functions}, apply 
the same reasoning as for the two plaquette lattice and arrive at the same 
conclusion: local Wilson loops cannot be used to distinguish between non-local 
states; one needs a Wilson loop that goes through both of the non-identity 
links in order to distinguish the two physical states.

\fig{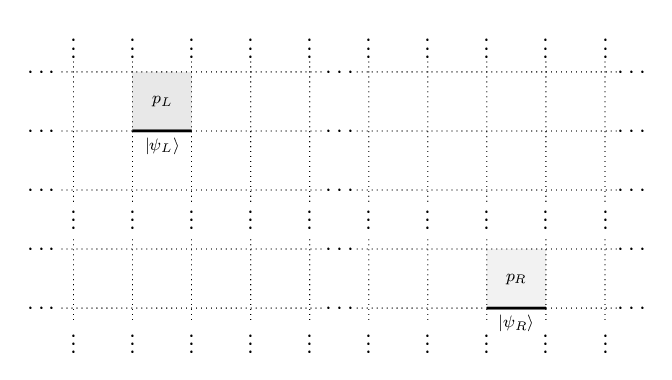}{0.7}{A geometrically inseparable state on a 
lattice with a non abelian gauge group. The state is such that all the links 
except the two labeled ones are set to the group identity.}

The difficulty of expressing states on lattices with the quaternion group is 
also apparent when we look at electric states. The quaternion group has five 
representations, but only four are one-dimensional, which is insufficient to 
construct a local electric basis for the physical states. This does not mean 
that the quaternion group gauge theory, and nonabelian theories in general, do 
not have electric states, but only that such states are not all expressible as 
a tensor product of electric states on individual links. Specifically, we can 
always create electric states by applying Wilson loop operators on the electric 
vacuum $\ket{0}$, which is the electric state corresponding to the trivial 
representation. However, doing so with extended loops in nonabelian theories 
can result in inseparable states. For example, we can apply a properly 
normalized version of the two-plaquette loop operator using the fixed lattice 
in \figref{figures/twop-phys-hs}:
\begin{align}
	2 U_{\dplaqd}^0 \ket{0} 
		&= tr(U_L U_R) \dfrac{1}{|G|}\sum_{g_L, g_R \in G}\ket{g_L g_R} \nonumber\\
		&= \dfrac{1}{8}\Big(\ket{+I}\ket{+I} + \ket{-I}\ket{-I} - \ket{+I}\ket{-I} - \ket{-I}\ket{+I} \nonumber\\
		&+ \sum_{a \in \mset{1, 2, 3}}
			\ket{+i\sigma_a}\ket{-i\sigma_a} + \ket{-i\sigma_a}\ket{+i\sigma_a} - 
				\ket{+i\sigma_a}\ket{+i\sigma_a} - \ket{-i\sigma_a}\ket{-i\sigma_a} \Big),
\end{align}
which is a state that cannot be factorized. By contrast, for a $Z_2$ abelian 
group, one would obtain:
\begin{align}
	U_{\dplaqd}^0 \ket{0} 
		&= tr(U_L U_R) \dfrac{1}{2}\Big(\ket{\SR}\ket{\SR} + \ket{\SR}\ket{\SL} 
			+ \ket{\SL}\ket{\SR} + \ket{\SL}\ket{\SL}\Big) \nonumber\\
		&= \dfrac{1}{2}\Big(\ket{\SR}\ket{\SR} - \ket{\SR}\ket{\SL} 
			- \ket{\SL}\ket{\SR} + \ket{\SL}\ket{\SL}\Big) \nonumber\\
		&= \dfrac{1}{\sqrt{2}}\Big(\ket{\SR} - \ket{\SL}\Big) \otimes
			\dfrac{1}{\sqrt{2}}\Big(\ket{\SR} - \ket{\SL}\Big),
\end{align}
where the states $\ket{\SR}$ and $\ket{\SL}$ are eigenstates of $U$ with 
eigenvalues $+1$ and $-1$, respectively.

In principle, one could attempt to fully fix the gauge in the field basis by 
constraining the ancillary global gauge transformation. In order to do so, one 
must impose some condition on the gauge fields. Looking at the states in 
\eqref{qg-extended-state-functions}, the two terms in each state are related by 
a global gauge transformation and therefore part of the same gauge orbit. 
However, both states are symmetric under a $(L) \leftrightarrow (R)$ exchange, 
and one can conclude that a gauge fixing condition cannot be local, since a 
local condition, applicable equally to both factors, cannot select a single 
term in both states. One is left with non local conditions. Furthermore, even 
non local conditions that are invariant to link exchanges, such as conditions 
of the form $\max_{g} \sum_{x, \mu} O[g u_{x, \mu} g^{-1}]$ which maximize some 
functional that only has local terms, $O$, over all global gauge 
transformations\cite{PhysRevD.41.2586} can fail to fix the gauge, since they 
would also fail to select a single term in either of the states in 
\eqref{qg-extended-state-functions}. These difficulties are reminiscent of the
Gribov ambiguity from the continuum.

\subsection{$SU(N)$}

For continuous groups, the dimension of the local Hilbert space (physical or 
unphysical) is not finite. This makes the exact form of counting arguments that 
were used for discrete groups impossible to use. Instead, the natural extension 
to continuous groups is to attempt to construct a map from pairs of physical 
basis states on two regions of a lattice to physical basis states on the whole 
lattice. Or, conversely, show that such a construction is impossible because 
the whole lattice can support multiple physical states for a single pair of 
physical states on the smaller regions, when those regions are taken separately 
from the rest of the lattice. In other words, one can show that larger lattices 
support physically distinct states that are indistinguishable using local 
physical operators or products of local physical operators\footnote{It should 
be noted that the use of ``locality'' makes the argument somewhat weak. A 
physical operator can appear non-local when expressed in terms of unphysical 
operators (see \secref{overview}).}. Specifically, one can show that $\exists 
g_a, g_b^1, g_b^2 \in SU(N)$ with 
\begin{align}
	\eqlabel{extended-space-condition}
	g_b^1 &\cong g_b^2,\nonumber\\
    g_a g_b^1 &\ncong g_a g_b^2,
\end{align}
where ``$\cong$'' denotes class equivalence. The physical meaning of that 
statement is that one could take the two-plaquette lattice in 
\figref{figures/twop-phys-hs} and states
\begin{align}
\eqlabel{sun-extended-states}
\ket{\psi_1} = \int_{G}\diff u \ket{u g_a u^{-1}}_{\dplaql}\ket{u g_b^1 u^{-1}}_{\dplaqr},\nonumber\\
\ket{\psi_2} = \int_{G}\diff u \ket{u g_a u^{-1}}_{\dplaql}\ket{u g_b^2 u^{-1}}_{\dplaqr},
\end{align}
where we labeled the kets inside the integral as belonging to the left or 
right plaquette and $\diff u$ is the Haar measure. As mentioned before, gauge 
invariant operators diagonal in the field basis take the form $f_c(U_\MC)$, 
where $\MC$ is a closed curve of links on the lattice and $f_c$ is a class 
function. Assume $W_\MC$ is such an operator with some $f_c$ injective over 
conjugacy classes. That is, $f_c(g_1) = f_c(g_2)$ if and only if $g_1 \cong 
g_2$. Applying $W_{\dplaql}$ and $W_{\dplaqr}$ to the states $\ket{\psi_1}$ and 
$\ket{\psi_2}$ in \eqref{sun-extended-states}, we obtain:
\begin{align}
	W_{\dplaql}\ket{\psi_1} &= f_c(g_a)\ket{\psi_1},\nonumber\\
	W_{\dplaqr}\ket{\psi_1} &= f_c(g_b^1)\ket{\psi_1},\nonumber\\
	W_{\dplaql}\ket{\psi_2} &= f_c(g_a)\ket{\psi_2},\nonumber\\
	W_{\dplaqr}\ket{\psi_2} &= f_c(g_b^2)\ket{\psi_2}.
\end{align}
Since $f_c(g_b^1) = f_c(g_b^2)$, it follows that $W_{\dplaql}\ket{\psi_1} = 
W_{\dplaql}\ket{\psi_2}, W_{\dplaqr}\ket{\psi_1} = W_{\dplaqr}\ket{\psi_2}$, 
and the two states are indistinguishable using the $W_{\dplaql, \dplaqr}$ 
operators. However, $W_{\dplaqd}$, the operator associated with the curve that 
goes around both plaquettes (without crossing), when applied to the two states 
above yields:
\begin{align}
	W_{\dplaqd}\ket{\psi_1} &= f_c(g_a g_b^1)\ket{\psi_1},\nonumber\\
	W_{\dplaqd}\ket{\psi_2} &= f_c(g_a g_b^2)\ket{\psi_2},
\end{align}
which are different by the assumption $g_a g_b^1 \ncong g_a g_b^2$. The 
existence of group elements satisfying \eqref{extended-space-condition} can be 
shown explicitly for $SU(N)$:
\begin{align}
	g_a = g_b^1 &= 
		\begin{pmatrix}
			\w{e^{i\theta}}  &                  &       &        \\ 
                             & \w{e^{-i\theta}} &       &        \\ 
                             &                  & \w{1} &        \\ 
                             &                  &       & \ddots 
        \end{pmatrix} \nonumber\\
	g_b^2 &= 
		\begin{pmatrix}
			\w{e^{-i\theta}} &                 &       &        \\ 
			                 & \w{e^{i\theta}} &       &        \\ 
			                 &                 & \w{1} &        \\ 
			                 &                 &       & \ddots
	    \end{pmatrix},
\end{align}
with the dots being ones and $\theta \in (0, \pi)$. The two are conjugate since 
$g_b^2 = h g_b^1 h^{-1}$, with
\begin{align}
	h = \begin{pmatrix}
		\w{0}  & \w{1} & \w{} &      \\ 
		    -1 & 0     &      &      \\ 
		       &       &    1 &      \\ 
		       &       &      & \ddots
		\end{pmatrix}.
\end{align}
However, $g_a g_b^1$ and $g_a g_b^2$ are clearly not conjugate, since their 
characters in the fundamental representation are different:
\begin{align}
	\tr(g_a g_b^1) &= 
		\tr \begin{pmatrix}
			\w{e^{2i\theta}} & \w{}          & \w{} &       \\ 
			                 & e^{-2i\theta} &      &       \\ 
			                 &               & 1    &       \\ 
			                 &               &      & \ddots
			\end{pmatrix}
		= (N-2) + 2 \cos(2\theta) \nonumber\\
	\tr(g_a g_b^2) &= 
		\tr \begin{pmatrix}
			\w{1} & \w{} & \w{} &      \\ 
			      & 1    &      &      \\ 
			      &      & 1    &      \\ 
			      &      &      & \ddots
			\end{pmatrix}
		= N.
\end{align}

\newcommand*{\defeq}{\stackrel{\text{def}}{=}}
\newcommand{\sutwo}[4]{\begin{pmatrix}\w{#1} & \w{#2} \\ #3 & #4\end{pmatrix}}

Going further into a general treatment for $SU(N)$ is difficult, but one can 
gain more insight by restricting the discussion to $SU(2)$. Conjugacy classes 
in $SU(2)$ are fully described by the character of the fundamental 
representation. That implies that classes can be parameterized as $\Cl_\theta 
\defeq \Cl(\dg(\theta))$ with $\theta \in [0, \pi]$, where we define 
$\dg(\theta) =  \diag(e^{i\theta}, e^{-i\theta})$. Without loss of generality, 
we can pick $g_a = \dg(\theta_a), g_b^1 = \dg(\theta_b)$ in 
\eqref{extended-space-condition}. We then want to see what conjugacy class $g_a 
g_b^2$ belongs to as $g_b^2$ spans $\Cl(g_b^1)$. To do so, we pick an arbitrary 
$SU(2)$ element $h = \sutwo{\alpha}{\beta}{-\beta^*}{\alpha^*}$, with 
$|\alpha|^2 + |\beta|^2 = 1$ and calculate $\tr (g_a h g_b^1 h^\dagger)$ to 
obtain:
\begin{align}
\tr (g_a h g_b^1 h^\dagger) = 2 |\alpha|^2 \cos(\theta_a + \theta_b)  + 
	2 (1 - |\alpha|^2) \cos(\theta_a - \theta_b).
\end{align}
It follows that the the field basis of physical states on two plaquettes is 
described by three parameters: $\theta_a, \theta_b \in [0, \pi]$, and 
$|\alpha|^2 \equiv x \in [0, 1]$. The later corresponds to distinct physical 
states only if  $\cos(\theta_a + \theta_b) \ne \cos(\theta_a - \theta_b)$ which 
is equivalent to $\theta_a, \theta_b \in (0, \pi)$. This is equivalent to the 
statement that a basis for physical states on a two-plaquette $SU(2)$ lattice 
consist of simultaneous eigenstates of Wilson loop operators $U_{\dplaql}, 
U_{\dplaqr}, U_{\dplaqd}$, where $U_\MC$ are like $W_\MC$ above, but with $f_c 
= \tr$, the remaining operator, $U_{\dplaqx}$, being 
related~\cite{WATSON1994385} to the other three by the following $SU(2)$ 
Mandelstam constraint:
\begin{align}
U_{\dplaqx} = U_{\dplaql}U_{\dplaqr} - U_{\dplaqd}.
\end{align} 
We can, therefore, express the physical Hilbert space on a two-plaquette 
$SU(2)$ lattice as 
\begin{align}
	\eqlabel{nadphs}
	\HS\phys_{\dplaqd} = (P \otimes P) \oplus (P \otimes B) 
		\oplus (B \otimes P) \oplus (B \otimes B \otimes X),
\end{align}
where $P = \lspan \big\{\ket{\dg(0)}, \ket{\dg(\pi)}\big\}$, $B = \lspan 
\big\{\ket{\Cl_\theta} |\; \theta \in (0, \pi)\big\}$, and $X = 
\lspan\big\{\ket{x} |\; x \in [0, 1]\big\}$. This shows a similar structure to 
the physical space of the two plaquette lattice with a Quaternion group, but 
where $X$ enters as a factor in a term rather than being a term in a direct 
sum. For simplicity, in the discussion that follows, we will focus only on the 
last term of \eqref{nadphs}. It is, in principle, possible to write $X$ as a 
product of two factors in the sense that one may be able to find $\HS\phys_1$ 
and $\HS\phys_2$ such that there is a one-to-one mapping $(\theta_a, \theta_b, 
x) \overset{f}{\longleftrightarrow} (u_1, u_2)$, with field basis vectors 
$\ket{u_1} \in \HS\phys_1, \ket{u_2} \in \HS\phys_2$ and with $u_1, u_2$ taking 
values in some connected subspace of $\mathbb{R}$ such that $u_i$ have the 
appearance of fields. However, this cannot be done while preserving certain 
properties of the mapping. For example, if we wanted to analytically relate the 
algebras of $\HS\phys_i$ with the algebra of $\HS\phys_{\dplaqd}$, $f$ would 
need to be analytic and thus continuous. That is, we would need to find a 
bijective and continuous function from a connected subspace $A = \{(\theta_a, 
\theta_b, x)\} \subset \mathbb{R}^3$ to a connected subspace $B = \{(u_1, 
u_2)\} \subset \mathbb{R}^2$, which is impossible, as can be seen from the 
following argument: Consider a closed curve $\mathcal{C} \subset B$. Then $B 
\setminus \mathcal{C}$ is disconnected. By continuity of $f$, the pre-image $A 
\setminus \mathcal{S} = f^{-1}(B \setminus \mathcal{C})$, where $\mathcal{S} = 
f^{-1}(\mathcal{C})$, should also be disconnected, which implies that 
$\mathcal{S}$ is a surface in $A$. The surface $\mathcal{S}$ is connected since 
$\mathcal{C}$ is connected by construction. Let $x_1, x_2 \in \mathcal{S}$ be 
two distinct points. The set $\mathcal{S}' = \mathcal{S} \setminus \{x_1, 
x_2\}$ remains connected\footnote{It is possible for $\mathcal{S}$ to be a 
surface with singular points and then the statement would not always be true; 
however, one can always choose $x_1$ and $x_2$ such that they do not coincide 
with singular points on the surface.}. However, the image of $\mathcal{S}'$ is 
now $\mathcal{C}' = \mathcal{C} \setminus \{f(x_1), f(x_2)\}$ and the points 
$f(x_1)$ and $f(x_2)$ divide $\mathcal{C}$ into two disconnected curve 
segments. Thus, we arrive at a contradiction, since $\mathcal{S}'$ is 
connected, but its image $\mathcal{C}'$ through a continuous function is not.

We can also prove a less general but possibly more illuminating statement: 
there is no way to construct the function $f$ such that it does not mix the 
subspaces of $\theta_a$ and $\theta_b$. That is, if $(u, v) = f(\theta_a, 
\theta_b, x)$ and $f$ is bijective, there is no bijective $g$ such that
\begin{align}
	u &= g(\theta_a, x)\nonumber\\
	v &= g(\theta_b, x).
\end{align}
To see this, take $x_1 \ne x_2$ such that $(u_1, v_1) = f(\theta_a, \theta_b, 
x_1)$ and $(u_2, v_2) = f(\theta_a, \theta_b, x_2)$. Then $u_1 = g(\theta_a, 
x_1), v_2 = g(\theta_b, x_2)$. Since $(u_1, v_1), (u_2, v_2) \in \HS\phys_1 
\otimes \HS\phys_2$ then also $(u_1, v_2) \in \HS\phys_1 \otimes \HS\phys_2$. 
This implies that $\exists (\theta_a', \theta_b', x')$ such that $(u_1, v_2) = 
f(\theta_a', \theta_b', x')$. But that implies that $g(\theta_a, x_1) = 
g(\theta_a', x')$, which, by the bijective condition of $g$ implies that 
$\theta_a = \theta_a', x_1 = x'$. Similarly, we obtain that $\theta_b = 
\theta_b', x_2 = x'$ and thus $x_1 = x_2$, which is a contradiction.

While the existence of the condition in \eqref{extended-space-condition} 
exists for higher $SU(N)$, the one-to-one mapping between the character in the 
fundamental representation and the conjugacy class does not necessarily hold. 
For example, in $SU(4)$, $\diag(1, -1, 1, -1)$ and $\diag(i, -i, i, -i)$ have 
the same trace, but are not in the same conjugacy class. The resulting 
complexity is beyond the scope of this paper.

\section{The role of matter fields}
\label{sec:na-matter}

We now turn to theories where both matter and gauge fields are present. In 
contrast with pure gauge theories, the addition of matter fields results in the 
ability to factor the Hilbert space locally. For a somewhat related 
discussion, please see~\cite{harlow15worm}.

We use a similar strategy as with pure gauge lattices: fix the gauge in order 
to understand how the physical space looks like, then construct, from the 
unphysical algebra, gauge invariant operators that generate gauge invariant 
states which span a space isomorphic to the gauge fixed space. The arguments 
presented here hold in a general sense, being applicable to general groups and 
sets they act on. However, for simplicity, we restrict the discussion to matter 
fields that take the form of vectors.

We assume a matter field that transforms in the familiar way 
(\eqref{gt-matter-sector}) under gauge transformations, which we repeat here:
\begin{align}
	\eqlabel{gt-matter}
	\ket{\phi(x)} &\rightarrow 	\ket{\phi'(x)} = \ket{g(x) \phi(x)}.
\end{align}

We fix the gauge as before, by setting links to the group identity on a 
maximal tree. As before, we are left with a global gauge transformation, which 
leaves the identity links unchanged. However, matter fields at vertices 
attached to those links are not invariant under the global gauge 
transformation, and we cannot simply discard all but the unfixed links from the 
gauge-fixed space. We are left with \dof\ corresponding to the unfixed links as 
well as those corresponding to matter fields at the vertices. The global gauge 
transformation now amounts to an overall phase of the matter fields. We can fix 
the global gauge by rotating everything such that the field on one chosen 
vertex $x_0$ points in a gauge direction of our choice. Specifically, field 
basis vectors for the matter fields can be written as
\newcommand{\dket}[2]{\ket{#1}\otimes\ket{#2}}
\newcommand{\tphi}{\tilde{\phi}}
\begin{align}
	\eqlabel{matter-field-decomp}
	\ket{\phi(x)} &= \ket{v(x) 1_\phi \tilde{\phi}(x)},
\end{align}
with $v(x)$ being gauge group $G$ valued, $1_\phi$ being a representative 
vector that elements in $G$ act on faithfully, and $\tilde{\phi}(x)$ being a 
scalar. We assume, for simplicity, that the space of the kets in the 
RHS of \eqref{matter-field-decomp} can be factored as $\ket{v(x) 1_\phi 
\tilde{\phi}(x)} = \dket{v(x)}{\tphi(x)}$. We fix the gauge fully by picking a 
$x_0$ and using a global gauge transformation
\begin{align}
\dket{v(x)}{\tphi(x)} \rightarrow \dket{v(x_0)^{-1} v(x)}{\tphi(x)},
\end{align}
which implies
\begin{align}
\dket{v(x_0)}{\tphi(x_0)} \rightarrow \dket{I_G}{\tphi(x_0)},
\end{align}
where $I_G$ is the gauge group identity. We now have $N_V - 1$ group-valued 
\dof, where $N_V$ is the number of vertices in the lattice, corresponding to 
the gauge portion of the matter fields at all vertices except for $x_0$. From 
graph theory, we know that a maximal tree always has $N_V - 1$ edges, which 
implies that we are also left with $N_L - (N_V - 1)$ unfixed link \dof. The 
total number of gauge-valued \dof\ is then
\begin{align}
	N_G = N_V - 1 + N_L - (N_V - 1) = N_L.
\end{align}
That is, after full gauge fixing, we have precisely as many gauge-valued \dof\ 
as we have links in the lattice and as many scalar valued \dof\ as we have 
vertices. We can, therefore, construct a physical Hilbert space from the 
unphysical one by a re-definition of variables:
\begin{align}
	\eqlabel{new-dof}
	\phi'(x) &= \tphi(x) \nonumber\\
	u_i'(x) &= v(x)^{-1} u_i(x) v(x + \hat{i}),
\end{align}
where $x + \hat{i}$ denotes a neighbouring vertex in the $i$ direction from 
$x$. Both the redefined variables are now gauge invariant quantities. The 
primed matter field is invariant by construction. For $u'$, we note that the 
unphysical field $v$ transforms as $v(x) \rightarrow g(x)v(x)$. Then, using
\eqref{gt-gauge-sector}
\begin{align}
	\ket{u_i'(x)} &\rightarrow \ket{v(x)^{-1} g(x)^{-1} g(x) u_i(x) g^{-1}(x + \hat{i}) g(x + \hat{i}) v(x + \hat{i})} \nonumber\\
		&= \ket{v(x)^{-1} u_i(x) v(x + \hat{i})} \nonumber\\
		&= \ket{u_i'(x)}.
\end{align}

The physical interpretation of the fields $u'$ is that, in the corresponding 
electric basis, electric fluxes are now automatically associated with the local 
charges that create them. These charges are not coupled to local energy 
eigenstates of the primed matter sector, except possibly through dynamics.
The physical Hilbert space of the theory is then:
\begin{align}
	\eqlabel{factz-matter}
	\HSp = \bigotimes_x \HS_{\phi'} \otimes \bigotimes_{x, i} \HS_{u'}.
\end{align}

That is, the local fields $\phi'(x)$ and $u'_i(x)$ form a \emph{complete} set 
of gauge invariant \dof\ for the theory. It is important here that the 
matter field transform in the fundamental representation of the gauge group or 
a super-representation of it. If the matter field carries only a 
sub-representation of the gauge group, we can at most reduce the theory to a 
theory gauged with a subgroup $G'$ of $G$. This may still prove useful if 
$G'$ is abelian.

In terms of the the physical Hilbert space, it is important to note here that 
the space of a gauge-matter theory on a lattice is locally factorizable and 
exhibits none of the issues seen in pure gauge theories, abelian or otherwise. 
In fact, for the purpose of factorizing the Hilbert space of a theory into 
bipartite geometrical factors, a single matter field on the boundary between 
regions suffices~\cite{WATSON1994385}. To illustrate this, we note that the 
local factorization in \eqref{factz-matter} relies on the existence of entities 
that transform like standard matter fields, as in \eqref{gt-matter}. Such 
entities can be constructed from a matter field at a single point in space, 
$x_0$, and a Wilson line that connects $x_0$ with some other point, where 
Wilson lines are like Wilson loops (see \eqref{wilson-loop}) except on an open 
curve:
\begin{align}
	V_{\mathcal{C}}(x) 
		&= V(x_0) U_{\mathcal{C}}(x_0, x) \nonumber\\
		&= V(x_0) U_{i_0}(x_0) U_{i_1}(x_1) \cdots U_{i_n}(x_n)
\end{align}
where $x_k + \hat{i}_k = x_{k + 1}$ are points along the open curve 
$\mathcal{C}$ starting at $x_0$ and ending at $x$. One can then check that 
$V_{\mathcal{C}}(x)$ transforms the same as as $V(x)$ under a gauge 
transformation. This allows one to construct a physical space in a manner 
similar to that used to obtain the space in \eqref{factz-matter}. This, like 
any solution that adds \dof\ on a fictive boundary, is likely flawed. 
Specifically, such a solution would make it difficult to formulate a well 
defined area law for entanglement entropy. An area law is a dependence between 
the entanglement entropy of a specific state on the Hilbert space of a theory 
(typically the vacuum state) and the area of the boundary separating the two 
regions for which the entanglement entropy is calculated. To make such a law 
universal would imply that boundaries entering the calculation are entirely 
arbitrary. If \dof\ were defined only on boundaries, of which there were many, 
we would also have have multiple distinct Hilbert spaces, the choice of which 
would depend on precisely what regions we use in calculating the entanglement 
entropy. This would lead to an ill defined theory and an equally ill defined 
vacuum state.

\section{Conclusions}

We have shown that in some cases that could be reasonable analyzed, the 
physical Hilbert space of pure non-abelian lattice gauge theories does not 
admit a geometric factorization. We did so using simple counting arguments, 
since many other approaches can be complex and can drift towards fundamental 
issues in quantum mechanics and field theory that are not entirely settled.

We have also shown that the addition of matter fields changes the problem in a 
fundamental way: it makes a factorization both straightforward and universal 
across gauge groups. It leads to a well defined and complete set of physical 
\dof\ on Hamiltonian lattice gauge theories. It is perhaps fitting that reality 
appears to favor theories with matter.

The discussion here is somewhat narrow. We assume that the Hilbert space of 
matter fields is some space of square integrable functions. While this has to 
be true, for the simple reason that it induces a norm, which is needed for a 
sensible quantum theory, the precise Hilbert space will also depend on the 
potential in the Hamiltonian. Furthermore, when also involving interactions 
with a gauge field, the problem becomes more complex due to the interaction 
terms. The amount of information that can be gained from an analysis that does 
not involve dynamics for such cases is necessarily limited.

\bibliographystyle{apsrev4-1}
\bibliography{na}
\end{document}